%% file: bare_jrnl.tex
\documentclass[conference]{IEEEtran}
\IEEEoverridecommandlockouts

\ifCLASSINFOpdf
  \usepackage[pdftex]{graphicx}
\else  
   \usepackage[dvips]{graphicx}
\fi

\usepackage{amsmath,amsfonts}
\usepackage{amsthm}
\usepackage{algorithm}
\usepackage{array}
\usepackage[caption=true,font=normalsize,labelfont=sf,textfont=sf]{subfig}
\usepackage{textcomp}
\usepackage{stfloats}
\usepackage{url}
\usepackage{verbatim}
\usepackage{graphicx}
\usepackage{cite}
\usepackage{cuted}
\usepackage{setspace}
\usepackage{amssymb}
\usepackage{mathtools}
\usepackage{amsthm}
\usepackage{algpseudocode}
\usepackage{makecell}
\usepackage{caption}
\usepackage{float}
\usepackage{adjustbox}
\usepackage[table,x11names]{xcolor}
\usepackage{colortbl}
\usepackage{fancyhdr}
\usepackage{optidef}
\usepackage{siunitx}
\usepackage{stfloats}
\usepackage{color}
\usepackage{tikz}
\usepackage{pgfplots}

\newtheorem{theorem}{Lemma}

\newcommand\at[2]{\left.#1\right|_{#2}}

\pgfplotsset{compat=1.18}

\def\BibTeX{{\rm B\kern-.05em{\sc i\kern-.025em b}\kern-.08em
    T\kern-.1667em\lower.7ex\hbox{E}\kern-.125emX}}

\begin{document}

\title{
Energy-Efficient Power Control in Single-User M-MIMO-OFDM System with PA Nonlinearity 

\thanks{The work of S. Marwaha was funded in part by BMDV 5G COMPASS project within InnoNT program under Grant 19OI22017A. Research of P. Kryszkiewicz was funded by the Polish National Science Center, project no. 2021/41/B/ST7/00136. For the purpose of Open Access, the author has applied a CC-BY public copyright license to any Author Accepted Manuscript (AAM) version arising from this submission. The work of E. Jorswieck was supported partly by the Federal Ministry of Education and Research (BMBF), Germany, through the Program of Souverän, Digital, and Vernetzt Joint Project 6G-RIC under Grant 16KISK031.
}
}

\author{
\IEEEauthorblockN{Siddarth Marwaha and Eduard A. Jorswieck}
\IEEEauthorblockA{\textit{Institute of Communications Technology} \\
\textit{Technische Universität Braunschweig}\\
Braunschweig, Germany \\
\{s.marwaha, e.jorswieck\}@tu-braunschweig.de}
\and
\IEEEauthorblockN{Pawel~Kryszkiewicz }
\IEEEauthorblockA{\textit{Institute of Radiocommunications} \\
\textit{Poznan University of Technology}\\
Poznan, Poland\\
pawel.kryszkiewicz@put.poznan.pl}     
}

\maketitle

\begin{abstract}
Although multiple works have proposed energy-efficient resource allocation schemes for Massive Multiple-Input Multiple-Output (M-MIMO) system, most approaches overlook the potential of optimizing Power Amplifier (PA) transmission power while accounting for non-linear distortion effects.
Furthermore, most M-MIMO studies assume narrow-band transmission, neglecting subcarrier intermodulations at the non-linear PA for an Orthogonal Frequency Division Multiplexing (OFDM) system. Therefore, this work investigates the energy-efficient power allocation for a single-user equipment (UE) M-MIMO downlink (DL) system employing OFDM with nonlinear PAs. 
Unlike prior works, we model wide-band transmission using a soft-limiter PA model and derive a closed-form expression for the signal-to-distortion-and-noise ratio (SNDR) under Rayleigh fading and Maximal Ratio Transmission (MRT) precoding. Next, the Energy Efficiency (EE) function is defined considering two PA architectures and a distorted OFDM signal. 
We then propose a low complexity root-finding algorithm to maximize EE by transmit power adjustment.
Simulation results demonstrate significant EE gains over a fixed PA back-off baseline, with over $100\%$ improvement under both low and high path loss. Our findings reveal how the optimal operating point depends on the antenna count, the PA model, and the propagation conditions.

\end{abstract}


\IEEEpeerreviewmaketitle

\section{Introduction}
\label{sec:intro}
Massive multiple input multiple output (M-MIMO) technology has become a pervasive element of 5G systems, enabling high data rates. However, it comes at the cost of substantially higher energy consumption.
Therefore, significant effort is being made to increase its energy efficiency (EE)  \cite{3GPP38864}. Multiple works have proposed energy efficient resource, such as the number of antennas, users (UEs), and transmit power, allocation techniques\cite{9678321}. However, most of the reported works do not account for the impact of hardware-induced non-linearities, particularly those stemming from power amplifier (PA) imperfections. The operation of PAs in their non-linear region might be energy efficient, though the presence of non-linear distortion must be taken into account\textemdash also acknowledged by 3GPP \cite{3GPP38864} recently. 
In this context, the authors in \cite{bjornson2014massive} and \cite{9226127} incorporate hardware imperfections in the EE framework, with the aim of maximizing the EE of an M-MIMO system. However, both works assume a narrow-band transmission, not accounting for subcarrier intermodulation in an Orthogonal Frequency Division Multiplexing (OFDM) system. Although this assumption simplifies the analysis, it does not fully capture the practical behavior of the system. Moreover, as discussed in \cite{bjornson2014massive} (Sec. VIIB), modeling the distortion power as linearly proportional to the desired signal power is a local approximation valid only within the small power operating range. These assumptions do not allow for optimizing PA operation in its full range, for example, operation close to the saturation (clipping) region, which is the most interesting from the EE perspective. 

Thus, in this work, we maximize the EE of an M-MIMO OFDM base station (BS) serving a single UE in the downlink (DL) using Maximum Ratio Transmission (MRT) precoding over all available subcarriers, under the assumption of an independent and identically distributed (i.i.d.) Rayleigh fading channel. We adopt a soft-limiter PA model, which is known to be optimal from the Signal-to-Distortion Ratio (SDR) perspective \cite{raich2005_optimal_nonlinearity} and can effectively represent a PA preceded by a digital predistorter. These characteristics enable the analytical formulation of the Signal-to-Noise-and-Distortion power Ratio (SNDR). While our previous work \cite{marwaha2025optimaldistortionawaremultiuserpower} derived the SNDR expression, it addressed sum-rate maximization under a multi-UE Zero Forcing precoder, leading to a substantially different problem formulation and solution. In contrast, here we focus on EE maximization, employing a realistic PA power consumption model that accounts for transmitted signal distortion. Accordingly, the system model, analytical EE expression, and problem formulation are detailed in Sec. \ref{sec:sys_mod}. Our proposed low-complexity solution, which is guaranteed to converge to a stationary point of the non-convex optimization problem, is presented in Sec. \ref{sec:sol}. Finally, the numerical results and conclusions are discussed in Secs. \ref{sec:res} and \ref{sec:conc}, respectively.

\section{System Model}
\label{sec:sys_mod}
\begin{figure}
 \centering
 \resizebox{0.8\columnwidth}{!}{\input{system_model}}
 \caption{System model: $M$ transmit chains, including common digital signal processing (DSP), digital-to-analog converter (DAC), and up conversion, serving a single UE with $\beta$ mean channel gain and per-antenna allocated power of $P/M$.}
 \label{fig:model}
 \end{figure}
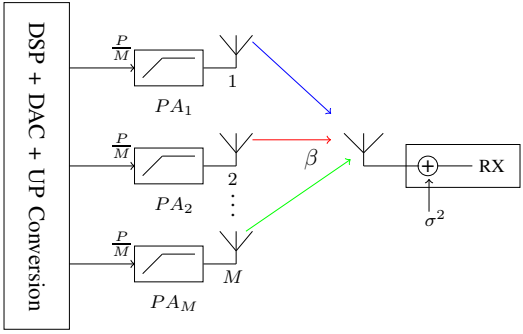  
We consider an $M$-antenna MIMO-OFDM BS serving a single, single-antenna UE as shown in Fig. \ref{fig:model}. A fully-digital MIMO implementation is considered utilizing $M$ transmit chains, each equipped with an $N$-point Inverse Fast Fourier Transform (IFFT) block. In each symbol $N_{\mathrm{U}}\leq N$ sub-carriers are modulated with complex symbols, resulting from per-antenna precoding of Quadrature Amplitude Modulation or Phase Shift Keing symbols \cite{Wachowiak2023}. The IFFT-processed signal at the $t$-th time sample can be denoted as $y_{m,t}$, where $t\in \{-N_{\mathrm{CP}},...,N-1\}$, $N_{\mathrm{CP}}$ denotes cyclic prefix duration in samples, and $m$ represents antenna index. 
The total mean power, averaged over time and across all front-ends, is
\begin{equation}
\label{eq_UE_power}
    P=\sum_{m}\mathbb{E}\left[ \left| y_{m,t}\right|^2 \right].
\end{equation} 
The time-domain signals, separately for each radio chain, are fed into the PAs. Here, the digital-to-analog conversion and upconversion are not explicitly modeled, as the complex baseband model is used for the PA. While multiple behavioral models for PAs exists, the one that minimizes the non-linear distortion power for OFDM is the soft limiter \cite{raich2005_optimal_nonlinearity}. This can be interpreted as an effective characteristic of the combination of a PA and predistorter, which is commonly used to reduce non-linear distortion \cite{Joung_2015_survey_EE_PA}. 
The output of the soft-limiter PA model for input $y_{m,t}$ 
\begin{equation}
    \hat{y}_{m,t} = \begin{cases} y_{m,t} & \mathrm{for} \left|y_{m,t}\right|^2 \leq P_{\mathrm{max}} \\
                  \sqrt{P_{\mathrm{max}}}e^{j \arg{(y_{m,t})}} & \mathrm{for} \left|y_{m,t}\right|^2 > P_{\mathrm{max}} \end{cases},
                  \label{eq_softlimiter}
\end{equation}
where $P_{\mathrm{max}}$ is the saturation power of each PA, and $\arg(~)$ denotes the argument of a complex number. For simplicity, the small-signal gain of this PA equals 1, though non-unitary gain can be embedded, e.g., by scaling $y_{m,t}$ at the PA input. The operating point of a PA can be described by the so-called Input Back-Off (IBO), defined as 
\begin{equation}
    \Psi =\frac{P_{\mathrm{max}}}{P/M},
    \label{eq_IBO_def}
\end{equation}
where the mean power at the PA input equals $P/M$ as a result of i.i.d. Rayleigh channel, MRT precoding and $N_{\mathrm{U}}\gg1$. While the OFDM waveform is transmitted through each frontend, allowing for $y_{m,t}$ modeling by complex-Gaussian distribution \cite{Wei_2010_dist_OFDM}, 
Bussgang decomposition of the PA output results in
\begin{equation}
\label{eq:bussgang_decomposition}
\hat{y}_{m,t} = \sqrt{\lambda} y_{m,t} + \bar{d}_{m,t}.
\end{equation}
where $\sqrt{\lambda}$ is the wanted signal scaling factor and $\bar{d}_{m,t}$ is the distortion signal, which is uncorrelated with the wanted signal $y_{m,t}$. Since both, the wanted signal distribution and PA characteristic (\ref{eq_softlimiter}) are known, the scaling coefficient equals    
\cite{kryszkiewicz2023efficiency}:
\begin{equation}
    \lambda\!=\!\!\left(\!\!\frac{
\mathbb{E} \left[  \hat{y}_{m,t}y_{m,t}^{*} \right]
    }{\mathbb{E} \left[ \left| y_{m,t} \right|^2\right]}\! \right)^2\!\!\!\!\!=\!\left(\!1\!-\!e^{-\Psi}\!\!+\!\!\frac{1}{2}\sqrt{\pi \Psi} \mathrm{\textrm{erfc}}\left(\sqrt{\Psi}\right)\!\right)^2\!.
    \label{eq_lambda}
\end{equation}
Observe that $\lambda$ depends only on $\Psi$. Moreover, it can be shown that $\lambda \in (0,1)$. Similarly, the mean distortion power is
\begin{equation}
\mathbb{E} \left[ \left| \bar{d}_{m,t} \right|^2\right]=\left(1-e^{-\Psi}-\lambda \right)\frac{P}{M}.
\label{eq_distortion_SISO}
\end{equation}
We are interested in the wanted signal power and distortion power at the receiver, combining signals received through $M$ wireless channels. According to the Bussgang decomposition, the wanted signal is scaled by $\sqrt{\lambda}$, which is equal across front-ends. Furthermore, since the M-MIMO gain for $M$ antennas serving a single UE is $M$ \cite{massivemimobook}, the received wanted signal power equals 
\begin{equation}
    S=\beta\lambda M P,
\end{equation}
where $\beta$ denotes large-scale fading between the transmitter and the receiver. It should be noted that omission of frequency selectivity is possible because for a sufficiently large number of utilized antennas, \emph{channel hardening} can be observed \cite{massivemimobook}. In addition, due to the nature of the OFDM transmission, the averaging over $N_{\mathrm{U}}$ frequency-selective channels will occur. 

As for the distortion power, its combining at the receiver depends on the wireless channel properties and the employed precoder \cite{mollen_spatial_char,Wachowiak2023}. 
Fortunately, for the i.i.d. Rayleigh channel considered in this paper, the distortion signals from each front-end are uncorrelated, resulting in an omnidirectional emission pattern. Therefore, each distortion signal of power $\mathbb{E} \left[ \left| \bar{d}_{m,t} \right|^2\right]$ passes through a wireless channel of large-scale fading coefficient $\beta$ adding up to the distortion from the remaining $M-1$ distortion signals, giving
\begin{equation}
    D=\eta \beta  \sum_{m}\mathbb{E} \left[ \left| \bar{d}_{m,t} \right|^2\right]=\eta \beta \left(1-e^{-\Psi}-\lambda \right)P,
\end{equation}
where $\eta$ is a scaling factor accounting for the leakage of some non-linear signal into the out-of-band frequency region. In \cite{lee2014characterization}, it has been estimated that around $\eta=2/3$ of the total distortion power falls in-band of utilized OFDM subcarriers.

Knowing that white noise with power $\sigma^2$ is added at the receiver over the occupied subcarriers, the resulting SNDR equals
\begin{equation}
    \gamma =  \frac{\beta \lambda M P}{\sigma^2 + \eta \beta \left(1-e^{-\Psi}-\lambda \right)P }.
\label{eq: SINR_MRT}
\end{equation}
A more comprehensive derivation of $\gamma$ for multi-UE system can be found in \cite{marwaha2025optimaldistortionawaremultiuserpower}.
Finally, the link capacity, treating the non-linear distortion as a noise-like signal, though this may not be true if advanced reception is performed \cite{Wachowiak2023}, is calculated as  
\begin{equation}
    R = N_{\mathrm{U}} \Delta f \log_2 (1 + \gamma), 
\label{eq: datarate2}
\end{equation}
where $\Delta f$ is OFDM subcarrier spacing. 

The total BS power consumption $P_{\mathrm{tot}}$ is defined as \cite{Marwaha_2023} 
\begin{eqnarray}
    P_{\mathrm{tot}} = P_{\mathrm{PA}}+P_{const} + M P_{\mathrm{SPRF}},   
\label{eq: egy_cons2}
\end{eqnarray}
where $P_{\mathrm{PA}}$ is the power consumed by the power amplifier, $P_{\mathrm{const}}$ accounts for constant power consumption, e.g., site-cooling, local oscillators, and $P_{\mathrm{SPRF}}$ accounts for the per RF chain power consumption, e.g., mixers.
Although we assume that the PA can clip some of the transmitted OFDM signal, this should, via the IBO value $\Psi$, influence $P_{\mathrm{PA}}$. Although there are multiple PA architectures \cite{ochiai2013analysis}, we consider a typical \textbf{Class B PA} for which, assuming soft-limiter behavioral PA modeling and OFDM signal, the consumed power summed over all $M$ transmit amplifiers equals \cite{kryszkiewicz2023efficiency,ochiai2013analysis}
\begin{equation}
     P_{\mathrm{PA}} =
    \frac{2MP_{\mathrm{max}}}{\sqrt{\pi \Psi}}\mathrm{erf}(\sqrt{\Psi}).
\label{eq_P_PA_B}
\end{equation}

The same component for a \textbf{Perfect PA}, i.e., for which consumed power equals emitted power, can be defined as,
\begin{eqnarray}
    P_{\mathrm{PA}} & = &  
    \frac{MP_{\mathrm{max}}}{\Psi} (1- e^{-\Psi}).
\label{eq_P_PA_perfect}
\end{eqnarray}

In the following, we are interested in observing the impact of PA non-linearity on the EE, where we aim to maximize the EE by optimizing the allocated power, solving
\begin{align}
\max_{P\geq 0}   &\quad  EE = \frac{R}{P_{tot}}.
\label{eq:optprob}
\end{align}

\section { Solution of the optimization problem}
\label{sec:sol}

It can be observed that the optimization problem in \eqref{eq:optprob} belongs to the family of fractional programming problem, making it suitable for transformation via Dinkelbach's method. However, the non-convexity
of \eqref{eq: datarate2} requires a hybrid approach combining Dinkelbach's method with successive convex approximation (SCA)\cite{zappone2014energy}, this would require iterative schemes that may increase computational complexity. 
Instead, we reformulate the problem as a scalar root-finding task that directly identifies a stationary point $\tilde{P}$ by solving 
\begin{equation}    
\at{\frac{\partial {EE}}{\partial P}}{P=\tilde{P}}=0.
    \label{eq_EE_d_Rk}
\end{equation}
Using properties of the $\frac{\partial {EE}}{\partial P}$ function, we employ the bisection search with function-specific initial point searching procedure, which guarantees convergence to the stationary point. 

First, to find $\tilde{P}$, we first analyze the properties of $\frac{\partial {EE}}{\partial P}$ using the following lemma: 
\begin{theorem}\label{lem:root_find}
The function $\frac{\partial {EE}}{\partial P}$ has at least a single root equal to the root of the function
\begin{equation}
f(P)= \frac{1}{R}\frac{\partial {R}}{\partial P} - \frac{1}{P_{\mathrm{tot}}}\frac{\partial P_{\mathrm{tot}}}{\partial P}
\label{eq_f}
\end{equation}
in the entire range of $P$, i.e., $P \in [ 0, \infty )$. 
\end{theorem}

\begin{proof}
The proof of Lemma \ref{lem:root_find} is provided in Appendix \ref{sec:proof_lemma_1}, where the derivative $\frac{\partial {EE}}{\partial P}$ is computed and its characteristic is analyzed. Our analysis shows that $f(P) \to +\infty$ as $P \to 0$, whereas, $f(P)\to 0^-$ as $P\to+\infty$. Therefore, by Intermediate Value Theorem, there must exist at least one root. 
\end{proof}
The analysis in Appendix \ref{sec:proof_lemma_1} enables us to propose Algorithm \ref{alg:subp1} to find the root of \eqref{eq_f}, where $f(P)$ is evaluated using (\ref{eq:dEEdP}).
The asymptotic behavior of $f(P)$ ensures that for sufficiently low $P_{L}$, the derivative $f(P_{L})$ is positive, while for sufficiently high $P_{U}$, the derivative $f(P_{U})$ is negative. Two initial \emph{while} loops are employed to find such points. A bisection search is then applied to find the root $P_C \in [P_{L}, P_{U}]$ within an accuracy defined by a small positive value $\delta$,
i.e., the root belongs to the range $[P_{C}-\frac{\delta}{2}, P_{C}+\frac{\delta}{2}]$. The bisection guarantees convergence in a logarithmic number of steps, which is significantly lower than the typical approach \cite{zappone2014energy} that combines the Dinkelbach method (for a fractional problem solution) with the successive convex approximation (SCA) method (for a non-convex function solution), followed by an internal numerical method for root finding, e.g., using Newton or bisection algorithm.

\begin{algorithm}
\caption{Distortion Aware Power Allocation}\label{alg:subp1}
\begin{algorithmic}[1]
\State Initialize: $P_{\text{test}}$, $P_L \gets P_{\text{test}}$, $P_U \gets P_{\text{test}}$, $\delta$
\While{$\left.f(P)\right|_{P=P_U} > 0$}
        \State $P_U \gets 2P_U$
    \EndWhile
    \While{$\left.f(P)\right|_{P=P_L} < 0$}
        \State $P_L \gets 0.5P_L$
    \EndWhile
\While{$P_U - P_L > \delta$}
    \State $P_C \gets 0.5P_L + 0.5P_U$
    \If{$\left.f(P)\right|_{P=P_C} > 0$} 
        \State $P_L \gets P_C$
    \Else
        \State $P_U \gets P_C$
    \EndIf
\EndWhile
\State $P_C \gets 0.5P_L + 0.5P_U$
\end{algorithmic}
\end{algorithm}

\section{Results}
\label{sec:res}

Here, the proposed EE solution is evaluated for a single UE served by $M = 4$ or $M = 32$ antennas, along with the simulation parameters provided in Table \ref{tab: sys_para}. We compare our results with a reference scenario, where the IBO $\Psi = 6$ dB is fixed, ensuring that the SDR is high enough to guarantee the mean error vector magnitude (EVM) of $4.5\%$ required in $5$G New Radio for $256$-QAM constellation \cite{3gpp_38141}.
\begin{table}[!ht]
\renewcommand{\arraystretch}{1.2}
    \centering
    \begin{adjustbox}{width=1.0\columnwidth}
        \begin{tabular}{|c|c|c|c|}
        \hline
        \rowcolor[HTML]{9B9B9B}
         $N_{\mathrm{U}}$ & $\Delta f$ & $\eta$ & $(\sigma^2)_{dBm}$  \\ \hline
        $1200$ & $15$ kHz & $\frac{2}{3}$ & $-174 \frac{dBm}{Hz} + 10  \cdot \log_{10}(N_{\mathrm{U}} \cdot \Delta f)$ \\ \hline
        \rowcolor[HTML]{9B9B9B}
        $P_{\mathrm{const}}$ & $P_{\mathrm{SPRF}}$ & $P_{\textrm{max}}$ & $(\beta)_{\mathrm{dB}}$   \\ \hline
        $348$ W & $23$ W & $160$ W & $(60, 150)$  \\ \hline  
        \end{tabular}
    \end{adjustbox}
\caption{Simulation parameters, with $P_{\mathrm{const}}$ and $P_{\mathrm{SPRF}}$ derived from \cite{Marwaha_2023}.}
\label{tab: sys_para}
\end{table}
\begin{figure}[!t]
    \centering
    \includegraphics[width=0.9\columnwidth]{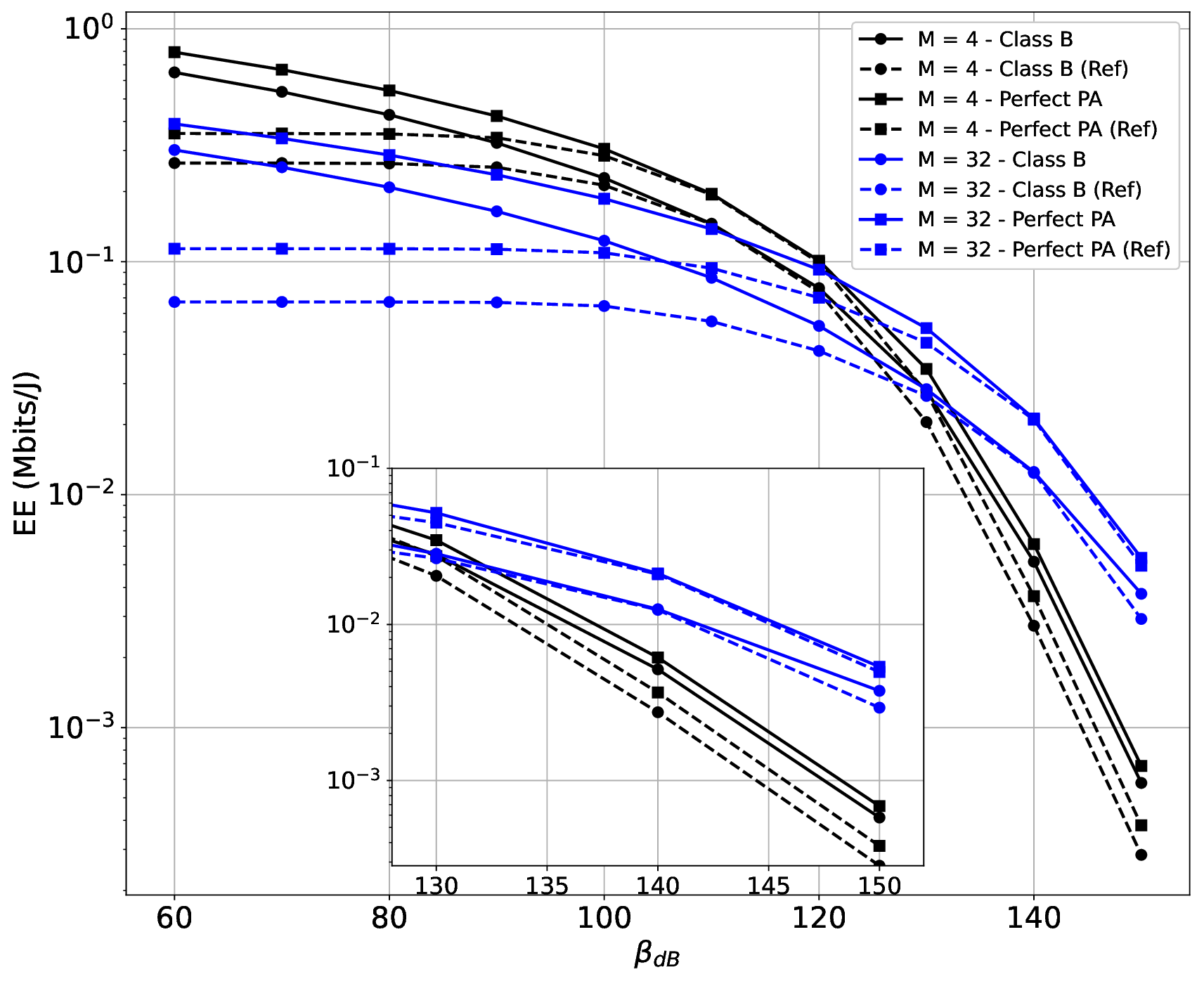}
    \caption{EE (Mbits/J) vs $\beta$ over different M.
    }
        \label{fig:EE}
\end{figure}
First, EE as a function of the path loss is shown in Fig. \ref{fig:EE}. As expected, EE drops with increasing path loss for all considered systems. It is visible that for any antenna-PA architecture configuration, the proposed solution for nearly all distances outperforms the reference solution. The proposed method yields the greatest gains at both low and high path loss levels,
exceeding $100\%$. For each PA architecture and number of antennas $M$, there exists a path loss value where the reference IBO $\Psi=6 dB$ is optimal, resulting in no improvement in EE.
As anticipated, the \textbf{Class B PA} has a lower EE than the \textbf{Perfect PA}. The difference is observed in the entire range of $\beta$ and varies from around $18\%$ to around $ 82\%$. Notably, the number of antennas $M$ has a significant impact on the EE. 
It is observable that for short distances, it is energy-efficient to use a lower number of antennas. Though as the path loss increases, the situation reverses and, e.g., for $\beta_{dB}>140 dB$ it is more energy-efficient to use $M=32$ antennas. 

\begin{figure}[!t]
    \centering
    \includegraphics[width=0.95\columnwidth]{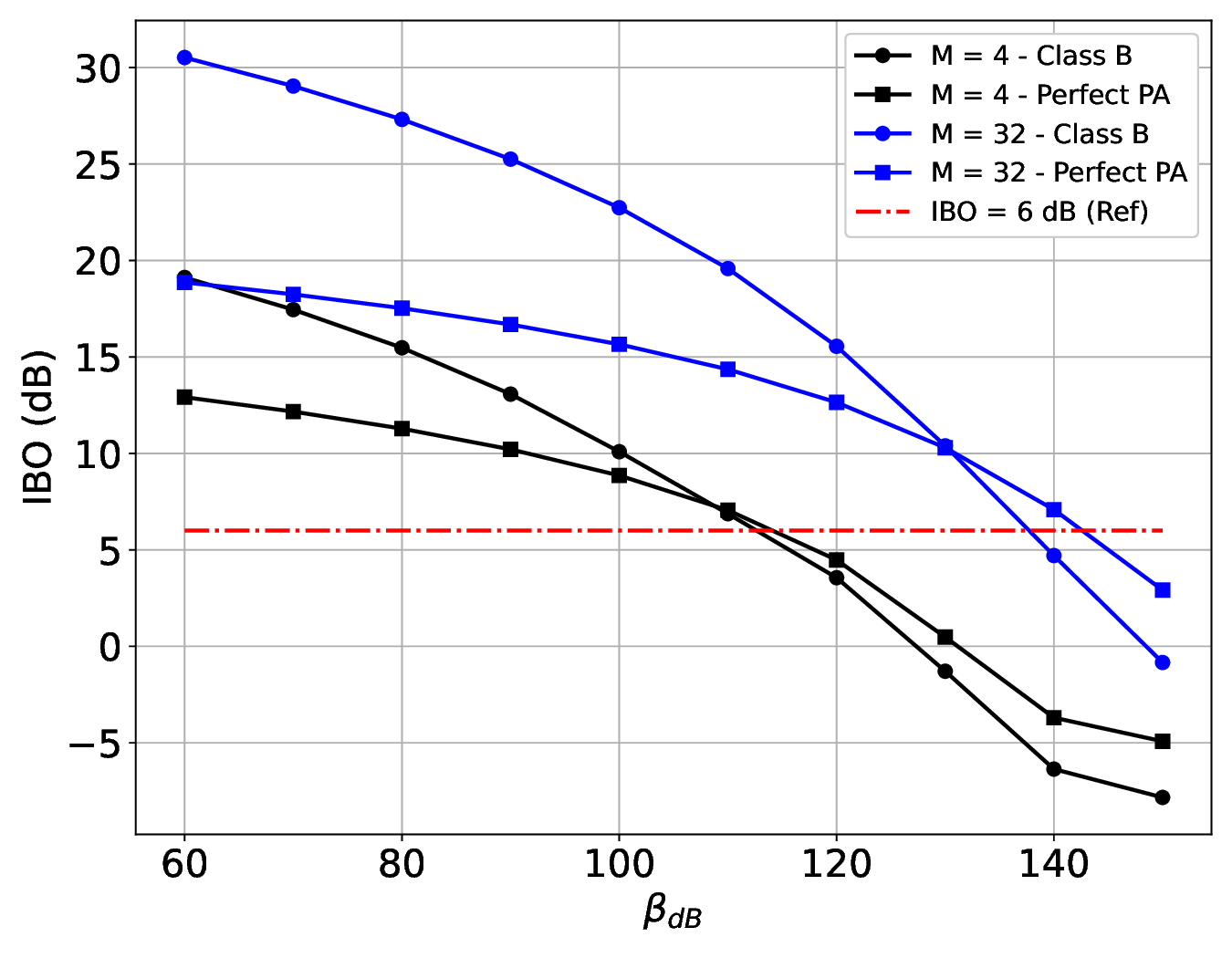}
    \caption{IBO (dB) vs $\beta$ over different M.}
    \label{fig:IBO}
\end{figure}

Next, Fig. \ref{fig:IBO} shows IBO as a function of the path loss. The proposed solution decreases IBO, or equivalently increases $P$, with increasing path loss in all the configurations. If a short link is considered, it is more energy-efficient to decrease the transmit power, significantly reducing non-linear distortion and the power consumed by the PA, relative to the reference solution. On the other hand, high path loss requires that stronger transmit power should be used. While this will result in increased distortion power, this is acceptable as the system's throughput is limited by the thermal noise $\sigma^2$. If the BS is equipped with a larger number of antennas, higher IBO is optimal due to the increased array gain. Finally, the optimal IBO depends on the selected PA architecture, and for the \textbf{Perfect PA} this it is lower than the optimal IBO for the \textbf{Class B PA} only for low path loss values. 

\section{Conclusion}
\label{sec:conc}

This work first analyzed the impact of nonlinear PA characteristics on the EE of a single-user M-MIMO OFDM system. Unlike prior studies, a wideband signal model was employed, with the soft-limiter used as the PA model. Subsequently, EE was maximized through transmit power adjustment. Notably, the proposed solution does not require an explicit transmit power constraint, as the distortion introduced by the PA inherently limits the result. Our distortion-aware power control achieves more than $100\%$ improvement in EE compared to fixed IBO baselines, under both low- and high-path loss regimes.
In addition, we observed that the optimal number of antennas maximizing EE depends on the UE–BS distance. For shorter distances, a smaller antenna array yields higher EE, whereas for larger distances, increasing the number of antennas becomes beneficial.

For future work, we aim to jointly optimize both the number of antennas and transmit power, potentially extending the analysis to multi-UE M-MIMO systems with inter-UE interference and imperfect Channel State Information (CSI). Furthermore, considering alternative PA models may provide additional insights. Finally, integrating receiver-side distortion mitigation techniques could further enhance EE in practical deployments.


\appendices

\section{Proof of Lemma \ref{sec:proof_lemma_1}}
\label{sec:proof_lemma_1}

The proof below employes the following limits \cite{abramowitz1964}
\begin{align}
    &\text{C1}: \lim_{x\to \infty} e^{-x} = 0, \text{C2}: \lim_{x \to \infty} \mathrm{erf}(x) = 1, \nonumber
    \\
    &\text{C3}: \lim_{x\to \infty} \mathrm{erfc}(x) = 0, \text{C4}: \lim_{x \to \infty} xe^{-x} = 0, \nonumber
    \\
    &\text{C5}: \lim_{x \to \infty}\frac{\mathrm{erfc}(x)}{x}=0,
    \label{eq:limits}
    \end{align}
and the following Taylor Series evaluated at $x=0$, formally refereed to as asymptotic expansions (denoted by $\sim$): 
\begin{align}
    &\text{A1}: \log_2(1+x) 
    \sim \frac{x}{\ln(2)}, \text{A2}: e^{-x} 
    \sim 1 - x + \frac{x^2}{2}, \nonumber
    \\
    &\text{A3}: \mathrm{erfc}(x) 
    \sim
    1 - \frac{2}{\sqrt{\pi}}x + \frac{2x^3}{3\sqrt{\pi}}.
    \label{eq:approxs}
    \end{align}
    
The partial derivative of $\frac{\partial {EE}}{\partial P}$ can be computed as
\begin{align}
\frac{\partial \mathrm{{EE}}}{\partial P} = 
\frac{\frac{\partial {R}}{\partial P} \cdot P_{\mathrm{tot}} - R \cdot  \frac{\partial P_{\mathrm{tot}}}{\partial P}}{P_{\mathrm{tot}}^2}.
\label{eq:ee_d}
\end{align}
Since $P_{\mathrm{tot}}$ is always positive, we focus only on the numerator of \eqref{eq:ee_d} and its root being equal to the root of a function  
\begin{equation}
    f(P) = \frac{1}{R}\frac{\partial {R}}{\partial P} - \frac{1}{P_{\mathrm{tot}}}\frac{\partial P_{\mathrm{tot}}}{\partial P}.
    \label{eq:der_EE}
\end{equation}
Let $B = N_{\mathrm{U}} \Delta f$, then $ \frac{\partial {R}}{\partial P}$ can be computed as 
\begin{align}
\label{eq_der_Rk}
    \frac{\partial {R}}{\partial P}=&
    \frac{BM \beta}{\ln(2) \left( 
    1+\frac{M\lambda P \beta}{\sigma^2+D}
    \right)
    }
    \frac{1}{\left(\sigma^2+ D\right)^2 }
    \nonumber
    \\&
    \cdot\left(
    \left(
    \frac{\partial \lambda}{\partial P}P
    +\lambda
    \right)
    \left( \sigma^2+ D \right)
    -\lambda P  \frac{\partial D}{\partial P}
    \right),
\end{align}
where 
\begin{align}
    \frac{\partial \lambda}{\partial P}=&
    -\frac{\Psi\sqrt{\lambda}}{P}
    \left( 
    e^{-\Psi}+\frac{1}{2}\sqrt{\frac{\pi}{\Psi}} \textrm{erfc} \left( \sqrt{\Psi} \right)
    \right),
\end{align}
and
\begin{align}
     \frac{\partial D}{\partial P}=&
     \eta \beta \left(1-  e^{-\Psi}-\lambda 
     -P \frac{\partial \lambda}{\partial P}-\Psi e^{-\Psi} \right),
\end{align}
recalling that IBO is defined as $\Psi=MP_{\mathrm{max}}/ P$. Furthermore, observe that $\frac{\partial P_{\mathrm{tot}}}{\partial P}$ can be computed as 
\begin{equation}
    \frac{\partial P_{\mathrm{tot}}}{\partial P} = \frac{\partial P_{\mathrm{PA}}}{\partial P},
\end{equation}
with 
\begin{equation}
    \frac{\partial P_{\mathrm{PA}}}{\partial P} =
    \begin{cases}
        \dfrac{\sqrt{\Psi}}{\sqrt{\pi}}\,\mathrm{erf}(\sqrt{\Psi}) - \dfrac{2}{\pi} \Psi e^{ -\Psi}, & \textbf{Class B PA} \\
        1 - e^{-\Psi} - \Psi e^{-\Psi}, & \textbf{Perfect PA}
    \end{cases}
\end{equation}
because the other components are independent of $P$. 
After substitution and some operations on \eqref{eq:der_EE}, $f(P)$ becomes
\begin{subequations}
    \begin{align}
   f(P) &= 
    \frac{B}{\log_2\left(1 + \frac{M \lambda  P \beta}{\sigma^2 + D} \right)} \frac{M\beta}{\ln(2) \left( 
    1+\frac{M\lambda P \beta}{\sigma^2+ D}
    \right)
    }  \label{eq:first_term}
    \\&
    \cdot \frac{\sqrt{\lambda}
    \left(1-e^{-\Psi}-\Psi e^{-\Psi} \right)}{\left(\sigma^2+ D\right)^2 } \label{eq:second_term}
    \\&
    \left( \sigma^2-\frac{\sqrt{\pi}}{2}\beta \eta M P_{max} \frac{erfc(\sqrt{\Psi})}{\sqrt{\Psi}}  \right) \label{eq:third_term}
    \\& 
     - 
    \frac{1}{P_{\mathrm{tot}}} \begin{cases}
        \dfrac{\sqrt{\Psi}}{\sqrt{\pi}}\,\mathrm{erf}(\sqrt{\Psi}) - \dfrac{2}{\pi} \Psi e^{ -\Psi}, & \textbf{Class B PA} \\
        1 - e^{-\Psi} - \Psi e^{-\Psi}, & \textbf{Perfect PA}
    \end{cases}  
    \label{eq:fourth_term}
\end{align}
\label{eq:dEEdP}
\end{subequations}
where we aim to solve $f(P) = 0$. In the following we show that \eqref{eq:dEEdP} has  at least a single root in the domain $P\geq0$. To find the root, we analyze the asymptotic behavior of $f(P)$ as $P \to 0$ and $P \to \infty$.

\textbf{Case 1 ($P \to 0$):}  Observe that as $P\to0$, $\Psi\to \infty$, and therefore using C1 and C3 $\lim_{\Psi\to \infty}\lambda = 1$  and $D = 0$. With $\lambda = 1$, $D = 0$ and $P = \frac{MP_{\mathrm{max}}}{\Psi}$, SNDR $\gamma$ can be written as 
\begin{equation}
    \gamma \sim \frac{M^2 P_{\mathrm{max}} \beta}{\sigma^2 \cdot \Psi}. 
\end{equation}
As $\Psi \to +\infty$, $\gamma \to 0$, and thus due to A1 \eqref{eq:first_term} simplifies to 
\begin{equation}
    \frac{BM\beta}{\ln(2)(1+\gamma)}\frac{1}{\log_2(1+\gamma)} \sim \frac{BM\beta}{\gamma} = \frac{B\sigma^2 \Psi}{M P_{\mathrm{max}}}.
\end{equation}
Next observe that in \eqref{eq:second_term} $\sqrt{\lambda} = 1$ and as $P \to 0$, using C1 and C4 $\lim_{\Psi\to \infty}\left(1-e^{-\Psi}-\Psi e^{-\Psi} \right) \to 1$. Therefore, \eqref{eq:second_term} results in $\frac{1}{(\sigma^2)^2}$. Furthermore, as $\Psi \to +\infty$, \eqref{eq:third_term} approaches $\sigma^2$ because of C5. Therefore, the product of the terms in \eqref{eq:first_term}, \eqref{eq:second_term}, and \eqref{eq:third_term} simplifies to $\frac{B\Psi}{MP_{\mathrm{max}}}$. 
Furthermore, substituting the asymptotic expansion of the error function \cite{abramowitz1964}, i.e. $x \to \infty$, 
\begin{equation}
    \mathrm{erf(x)} \sim 1 - \frac{e^{-x^2}}{x\sqrt{\pi}} \left(1 - \frac{1}{2x^2}\right)
    \label{eq:asym_erf}
\end{equation}
in $\frac{\partial P_{\mathrm{PA}}}{\partial P}$ for \textbf{Class B PA} and using C1 and C4 for \textbf{Perfect PA},  $\frac{\partial P_{\mathrm{PA}}}{\partial P}$ in \eqref{eq:fourth_term} can be approximated as    
\begin{align}
\frac{\partial P_{\mathrm{PA}}}{\partial P} \sim 
\begin{cases}
        \frac{\sqrt{\Psi}}{\sqrt{\pi}}, & \textbf{Class B PA} \\
         1 & \textbf{Perfect PA}.
    \end{cases}      
\end{align}
Moreover, note that in $P_{\mathrm{tot}}$, $P_{\mathrm{PA}}$ depends on $P$, while the other terms are independent. As $P\to0$ and $\Psi\to\infty$,  $P_{\mathrm{tot}}$ can be approximated as
\begin{align}
\begin{cases}
        P_{\mathrm{tot}} \sim P_{const} + M P_{SPRF}, & \textbf{Class B PA} \\
        P_{\mathrm{tot}} \sim P + P_{\mathrm{const}} + MP_{\mathrm{SPRF}}, & \textbf{Perfect PA}
    \end{cases}      
\end{align}
because using \eqref{eq:asym_erf} it can be easily shown that $P_{\mathrm{PA}} \to 0$ as $\Psi\to\infty$ for \textbf{Class B PA}, whereas, $P_{\mathrm{PA}} \to P$ for \textbf{Perfect PA} due to C1 and C4.  

Finally, using the approximations above, $f(P)$ simplifies to 
\begin{align}
f(P) &\sim
\begin{cases}
          \frac{B\Psi}{MP_{\mathrm{max}}} - \frac{1}{P_{\mathrm{tot}}} \left(\frac{\sqrt{\Psi}}{\sqrt{\pi}}\right), & \textbf{Class B PA} \\
        \frac{B\Psi}{MP_{\mathrm{max}}} - \frac{1}{P + P_{\mathrm{const}} + MP_{\mathrm{SPRF}}} , & \textbf{Perfect PA}
    \end{cases} \nonumber
    \\
    &\sim
    \begin{cases}
          \frac{1}{\sqrt{P}} \left(\frac{B}{\sqrt{P}}-\frac{\sqrt{M P_{\mathrm{max}}}}{\sqrt{\pi}P_{\mathrm{tot}}} \right), & \textbf{Class B PA} \\
        \frac{B}{P} - \frac{1}{P + P_{\mathrm{const}} + MP_{\mathrm{SPRF}}}, & \textbf{Perfect PA}
    \end{cases}       
\end{align}
showing that $f(P) \to +\infty$ as $P \to 0$.

\textbf{Case 2 ($P\to\infty$):} Observe that as $P\to\infty$, $\Psi\to 0$. As $\Psi\to 0$, using A2 and A3 yields 
\begin{align}
    \lambda & \sim 
    \left(\left(\Psi - \frac{\Psi^2}{2}\right)+ \frac{\sqrt{\pi \Psi}}{2} \left(1 - \frac{2\sqrt{\Psi}}{\sqrt{\pi}} + \frac{2\Psi^{2/3}}{3\sqrt{\pi}}\right)\right)^2 \nonumber\\ 
    &\sim \left(\Psi - \frac{\Psi^2}{2}+ \frac{\sqrt{\pi\Psi}}{2} - \Psi + \frac{\Psi^2}{3}\right)^2  
    \sim \frac{\pi\Psi}{4}
\end{align}
and \begin{align}
     D & \sim 
     \eta \beta \left(1 - (1-\Psi) -\frac{\pi\Psi}{4} \right)\frac{MP_{\mathrm{max}}}{\Psi} \nonumber \\ 
    &\sim \eta \beta (1 - \frac{\pi}{4})M P_{\mathrm{max}}. 
\end{align}
Let $C = \sigma^2 + \eta \beta (1 - \frac{\pi}{4})M P_{\mathrm{max}}$, then $\gamma = \frac{M\lambda P \beta}{C}$ can be written as 
\begin{equation}
    \gamma = \frac{M^2\lambda \beta P_{\mathrm{max}}}{C \cdot\Psi} = \frac{\pi M^2 \beta P_{\mathrm{max}}}{4C},
\label{eq_gamma_high_P}
\end{equation}
which is a constant value. Let $\gamma_0 = \gamma$ from (\ref{eq_gamma_high_P}), then \eqref{eq:first_term} can be written as
\begin{equation}
    \frac{M\beta}{\ln(2)(1+\gamma_0)}\frac{B}{\log_2(1+\gamma_0)} = \frac{BM\beta}{(1+\gamma_0) \ln(1+\gamma_0)},
\end{equation}
which is a constant as $\Psi \to 0$. Next observe that $\sqrt{\lambda} = \frac{\sqrt{\pi \Psi}}{2}$ and substituting A2 in $\left(1-e^{-\Psi}-\Psi e^{-\Psi} \right)$, it can be asymptotically expanded as $\Psi^2 - \frac{\Psi^3}{2}+ \dots \sim \Psi^2$. Similarly, substituting A3, the term in \eqref{eq:third_term} becomes    
\begin{align}
     \sigma^2\!-\!\frac{\sqrt{\pi}}{2}\beta \eta M P_{max} \! \left(\!\!\frac{1}{\sqrt{\Psi}} \!- \!\frac{2}{\sqrt{\pi}}\!\!\right) \! \sim \!
    \sigma^2\!-\!\frac{\sqrt{\pi}}{2}\beta \eta M P_{\mathrm{max}} \frac{1}{\sqrt{\Psi}}.
\end{align}
Next observe that the product of \eqref{eq:second_term} and \eqref{eq:third_term} can be expanded as  
\begin{align}
& \frac{\frac{\sqrt{\pi \, \Psi}}{2} \cdot \Psi^2}{C^2} \cdot \left( \sigma^2 - \frac{\sqrt{\pi}}{2} \beta \eta M P_{\mathrm{max}} \cdot \frac{1}{\sqrt{\Psi}} \right ) \nonumber \\
& \sim - \frac{\pi \beta \eta M P_{\mathrm{max}} \cdot \Psi^{5/2}}{4 C^2 \sqrt{\Psi}} 
 = - \frac{\pi \beta \eta M P_{\mathrm{max}} \cdot \Psi^2}{4 C^2}
\end{align}
because as $\Psi \to 0$, $\frac{1}{\sqrt{\Psi}}$ becomes large and therefore, $\sigma^2$ can be neglected. 
Finally, combining all the approximations, the product of \eqref{eq:first_term}, \eqref{eq:second_term} and \eqref{eq:third_term} results in 
\begin{equation}
   -\frac{\pi B \beta^2 \eta M^2 P_{\mathrm{max}} \Psi^2} {4 C^2 (1+ \gamma_0) \ln(1 + \gamma_0)}.
    \label{eq:final_exp_c2}
\end{equation}
Furthermore, observe that $\mathrm{erf}(\sqrt{\Psi}) = 1 - \mathrm{erfc}(\sqrt{\Psi})$. Substituting A2 and A3, $\frac{\partial P_{\mathrm{PA}}}{\partial P}$ in \eqref{eq:fourth_term} can be expanded as  
\begin{align}
\begin{cases}
           \frac{\partial P_{\mathrm{PA}}}{\partial P} 
           \sim
           \frac{2\Psi}{\pi}-\frac{2\Psi}{\pi} \sim 0, & \textbf{Class B PA} \\
        \frac{\partial P_{\mathrm{PA}}}{\partial P} 
        \sim
        \frac{x^2}{2} - \frac{x^3}{2} \sim 0 & \textbf{Perfect PA}
    \end{cases}   
\end{align}
and thus, the term in \eqref{eq:fourth_term} results in $0$. 
Finally, using the expansions above, $f(P)$ simplifies to 
\begin{equation}
    f(P) \sim -\frac{\pi B \beta^2 \eta M^2 P_{\mathrm{max}} \Psi^2} {4 C^2 (1+ \gamma_0) \ln(1 + \gamma_0)} 
    \label{eq:final_exp_c21}
\end{equation}
showing that $f(P)\to 0^-$ as $P\to+\infty$ and $\Psi\to 0$. 
Thus, by Intermediate Value Theorem, there must exist at least one root in $(0, +\infty)$ for both \textbf{Class B} and \textbf{Perfect PA}.


\bibliographystyle{IEEEtran}
\bibliography{IEEEabrv,bibtex/bib/IEEEexample}

\end{document}

%% file: system_model.tex
\pgfplotsset{compat=1.18}
\begin{tikzpicture}
    
    \node[rectangle, draw, minimum width=1cm, minimum height=5cm, align=center] (box) at (0,0) {};
    
    \node[rotate=270] at (0,0) {DSP + DAC + UP Conversion};

    \draw[->] (0.5, 1.5) -- ++(1, 0); 
    \node[rectangle, draw, minimum width=0.8cm, minimum height=0.4cm] at (2.02, 1.5) {
    \begin{tikzpicture}
        \draw[domain=2.7:3, samples=50, color=black] plot (\x, \x); 
        \draw[black] (3, 3) -- (3.5,3);
    \end{tikzpicture}}; 
    \draw[-] (2.55, 1.5) -- ++(0.5,0);
    \draw[-] (3.05, 1.5) -- (3.05, 2.0);
    \draw[-] (3.05, 1.7) -- (3.3, 2.0);
    \draw[-] (3.05, 1.7) -- (2.8, 2.0);

    \draw[->] (0.5, 0) -- ++(1, 0); 
    \node[rectangle, draw, minimum width=0.8cm, minimum height=0.4cm] at (2.02, 0) {
    \begin{tikzpicture}
        \draw[domain=2.7:3, samples=50, color=black] plot (\x, \x); 
        \draw[black] (3, 3) -- (3.5,3);
    \end{tikzpicture}}; 
    \draw[-] (2.55, 0) -- ++(0.5,0);
    \draw[-] (3.05, 0) -- (3.05, 0.5);
    \draw[-] (3.05, 0.2) -- (3.3, 0.5);
    \draw[-] (3.05, 0.2) -- (2.8, 0.5);

    \draw[->] (0.5, -1.5) -- ++(1, 0); 
    \node[rectangle, draw, minimum width=0.8cm, minimum height=0.4cm] at (2.02, -1.5) {
    \begin{tikzpicture}
        \draw[domain=2.7:3, samples=50, color=black] plot (\x, \x); 
        \draw[black] (3, 3) -- (3.5,3);
    \end{tikzpicture}}; 
    \draw[-] (2.55, -1.5) -- ++(0.5,0);
    \draw[-] (3.05, -1.5) -- (3.05, -1.0);
    \draw[-] (3.05, -1.3) -- (3.3, -1.0);
    \draw[-] (3.05, -1.3) -- (2.8, -1.0);

\node at (3.0, 1.3) {\fontsize{8}{12}\selectfont $1$};
\node at (2.1, 0.9) {\fontsize{8}{12}\selectfont $PA_1$};
\node at (1.3, 1.8) {\fontsize{8}{12}\selectfont $\frac{P}{M}$};

\node at (3.0, -0.2) {\fontsize{8}{12}\selectfont $2$};
\node at (2.1, -0.6) {\fontsize{8}{12}\selectfont $PA_2$};
\node at (1.3, 0.3) {\fontsize{8}{12}\selectfont $\frac{P}{M}$};

\node at (3.0, -0.5) {$\vdots$};

\node at (3.0, -1.7) {\fontsize{8}{12}\selectfont $M$};
\node at (2.1, -2.1) {\fontsize{8}{12}\selectfont $PA_M$};
\node at (1.3, -1.2) {\fontsize{8}{12}\selectfont $\frac{P}{M}$};

\draw[->, red](3.3, 0.4) -- (4.5, 0.4);
\draw[->, blue](3.3, 1.9) -- (4.5, 0.8);
\node at (4.2, 0.1) {\color{black}{$\beta$}};
\draw[->, green](3.2, -1.0) -- (4.8, 0.1);

    \begin{scope}[xshift=4cm] 

        \node[rectangle, draw, minimum width=0.4cm, minimum height=0.2cm] at (2.52, 0.00) {
            \begin{tikzpicture}
                \node at (2.52, 0) {+};
                \draw (2.52, 0) circle (0.15);
                \draw[-] (2.66, 0) -- (3.2, 0);
                \node at (3.5, 0) {\fontsize{8}{12}\selectfont RX};                              
            \end{tikzpicture}
        }; 
        \draw[<-] (2.0, -0.2) -- (2.0, -0.7);  
        \node at (2.1, -0.8) {\fontsize{8}{12}\selectfont $\sigma^2$};
        \draw[-] (1.0, 0.0) -- (1.85, 0.0);
        \draw[-] (1.0, 0.0) -- (1.0, 0.5);
        \draw[-] (1.0, 0.2) -- (1.3, 0.5);
        \draw[-] (1.0, 0.2) -- (0.7, 0.5);
        

    \end{scope}

\end{tikzpicture}
